\documentstyle[12pt]{article}
\def\be{\begin{equation}}
\def\ee{\end{equation}}
\def\bea{\begin{eqnarray}}
\def\eea{\end{eqnarray}}

\def\ra{\rightarrow}

\def\bar{\overline}

\def\a{\alpha}
\def\b{\beta}

\def\e{\epsilon}

\def\l{\lambda}

\def\bc{\begin{center}}
\def\ec{\end{center}}

\def\O{{\cal O}}
\def\PR#1#2#3{Phys. Rev.  {\bf #1}, (#3) #2}
\def\PRL#1#2#3{Phys. Rev. Lett. {\bf #1}, (#3) #2}
\def\PL#1#2#3{Phys. Lett. {\bf #1}, (#3) #2}

\def\NP#1#2#3{Nucl. Phys. {\bf #1}, (#3) #2}

\def\PTP#1#2#3{Prog. Theor. Phys. {\bf #1}, (#3) #2}

\begin{document}
\title{Quark Mixings in $SU(6)\times SU(2)_R$ \\
and Suppression of $V_{ub}$}
\author{
{M. Matsuda$^1$}\thanks{E-mail address:mmatsuda@auecc.aichi-edu.ac.jp}
{ and T. Matsuoka$^2$}\thanks{E-mail
address:matsuoka@kogakkan-u.ac.jp}
\\
\\
\\
{\small \it $^1$Department of Physics and Astronomy,
Aichi University of Education,}\\
{\small \it Kariya, Aichi, 448-8542 Japan}\\
{\small \it $^2$Kogakkan University, Nabari, Mie, 518-0498 Japan}}
\date{}
\maketitle
\vspace{-10.5cm}
\begin{flushright}
hep-ph/0003239\\
\end{flushright}
\vspace{10.5cm}
\vspace{-2.5cm}
\begin{abstract}
The quark mixing matrix $V_{CKM}$ is studied in depth 
on the basis of superstring inspired $SU(6)\times SU(2)_R$ 
model with global flavor symmetries. 
The sizable mixings between right-handed down-type quark 
$D^c$ and colored Higgs field $g^c$ potentially occur 
but no such mixings in up-type quark sector. 
In the model the hierarchical pattern of $V_{CKM}$ is 
understood systematically. 
It is shown that due to large $D^c$-$g^c$ mixings $V_{ub}$ 
is naturally suppressed compared to $V_{td}$. 
It is pointed out that the observed suppression of $V_{ub}$ 
is in favor of the presence of $SU(2)_R$ gauge symmetry 
but not in accord with generic $SU(5)$ GUT. 
\end{abstract}

{\sf keyword:CKM matrix, large mixing, unification model}
%%%%%%%%%%%%%%%%%%%%%%%%%%%%%%%%%%%%%%%%%%%%%%%%%%%%%%%%%%%%%%%%%%%%%
\newpage

%\section{Introduction}
Fermion masses and mixings are closely related to each other. 
Indeed, there have appeared many attempts to express 
the elements of Cabibbo-Kobayashi-Maskawa mixing matrix 
$V_{CKM}$ in terms of quark masses\cite{Q-mass}. 
The observed fermion masses have the hierarchical pattern 
though tiny mass pattern of neutrinos is still unclear. 
The $V_{CKM}$ also has peculiar pattern. 
It seems that these characteristic patterns shed some lights on 
the gauge symmetry and matter contents at the unification scale. 
The purpose of this work is to understand the following two 
challenging issues and then to explore the gauge symmetry 
and matter contents at the unification scale. 
\begin{enumerate}
\item
If we take a naive viewpoint of GUT, 
it is plausible that up- and down-type quarks reside in the same 
irreducible representation of GUT gauge group. 
This implies that the CKM matrix should be unit matrix 
in contrast to the experimental facts, 
which show nonzero values for off-diagonal elements\cite{PDG}. 
On the other hand, up- and down-type quarks have distinct 
hierarchical mass pattern each other($m_u/m_d<m_c/m_s<m_t/m_b)$. 
If Yukawa couplings of up- and down-quark sectors are independent 
each other, 
it is natural that the $V_{CKM}$ would have large off-diagonal 
elements. 
This is also inconsistent with experimental facts which show 
that $V_{CKM}$ might be almost unit matrix. 
How can we understand this property of the CKM matrix ? 

\item
Among the characteristic pattern of the CKM matrix, 
the asymmetric feature of the matrix is noticeable. 
Specifically, the element $V_{ub}$ in $V_{CKM}$ is rather small 
compared to $V_{td}$, i.e., 
%%%%%%%%%%%%%%%%%%%%%%%%%%%%%%%%%%%%%%%%%%%%%%%%%%%%%%%%%%
\bea
  |V_{td}| & \simeq & |V_{cd}\cdot V_{ts}| \simeq \l^3\,,
                            \qquad(\l \simeq 0.22) \nonumber\\
  |V_{ub}| & \simeq & \l \,|V_{cb}\cdot V_{us}| \simeq \l^4\,,
\eea
%%%%%%%%%%%%%%%%%%%%%%%%%%%%%%%%%%%%%%%%%%%%%%%%%%%%%%%%%%
where the second relation is suggested by the experimental results 
$|V_{ub}|/|V_{cb}|=0.08\pm0.02$\cite{PDG}. 
How can we understand this feature of the CKM matrix ? 
\end{enumerate}

\noindent
In the context of the string inspired $SU(6) \times SU(2)_R$ model 
with global flavor symmetries, 
it has been shown that the main pattern of fermion masses 
and mixings can be understood as a consequence of mixings between 
quarks(leptons) and extra particles\cite{Matsu1,Matsu2,Matsu3}. 
Along the previous works we investigate the structure of 
$V_{CKM}$ and quark masses in depth. 
In this paper we show that the above features can be naturally 
understood in the $SU(6) \times SU(2)_R$ model.

\vskip 1cm
%\section{Mass matrices and their diagonalization}
The model discussed here is the same as in 
Ref.\cite{Matsu1,Matsu2,Matsu3}. 
In this study we choose $SU(6) \times SU(2)_R$ 
as the unification gauge symmetry at the string scale $M_S$, 
which can be derived from the perturbative heterotic 
superstring theory via the flux breaking\cite{Matsu4}. 
In terms of $E_6$ we set matter superfields which consist of 
three family and one vector-like multiplet, i.e., 
%%%%%%%%%%%%%%%%%%%%%%%%%%%%%%%%%%%%%%%%%%%%%
\be
  3 \times {\bf 27}(\Phi_{1,2,3}) + 
        ({\bf 27}(\Phi_0)+\overline{\bf 27}({\bar \Phi})) .
\ee
%%%%%%%%%%%%%%%%%%%%%%%%%%%%%%%%%%%%%%%%%%%%%
Under $G= SU(6) \times SU(2)_R$, the superfields $\Phi$ in 
{\bf 27} of $E_6$ are decomposed into two groups as 
%%%%%%%%%%%%%%%%%%%%%%%%%%%%%%%%%%%%%%%%%%%%%%%%%%%%%%%%
\be
  \Phi({\bf 27})=\left\{
       \begin{array}{lll}
         \phi({\bf 15},{\bf 1})& : 
               & \quad \mbox{$Q,L,g,g^c,S$}, \\
          \psi(\overline{\bf 6},{\bf 2}) & : 
               & \quad \mbox{$(U^c,D^c),(N^c,E^c),(H_u,H_d)$}, 
       \end{array}
       \right.
\label{27}
\ee
%%%%%%%%%%%%%%%%%%%%%%%%%%%%%%%%%%%%%%%%%%%%%%%%%%%%%%%%
where $g$, $g^c$ and $H_u$, $H_d$ represent colored Higgs and 
doublet Higgs fields, respectively. 
Under $G$, doublet Higgs and 
color-triplet Higgs fields belong to  different representations and  
this situation is favorable to solve the triplet-doublet 
splitting problem. 
$N^c$ is the right-handed neutrino superfield and 
$S$ is an $SO(10)$-singlet. 
Although $D^c$ and $g^c$($L$ and $H_d$) 
have the same quantum numbers under the standard model 
gauge group $G_{SM} = SU(3)_c \times SU(2)_L \times U(1)_Y$, 
they belong to different irreducible representations of $G$. 
We assign odd $R$-parity for $\Phi_{1,2,3}$ and even for $\Phi_0$ 
and $\overline{\Phi}$, respectively. 
Since ordinary Higgs doublets have even $R$-parity, 
they belong to $\Phi_0$. 
It is assumed that $R$-parity remains unbroken down to 
the electroweak scale.

The gauge symmetry $G$ is spontaneously broken in two steps 
at the scale $\langle S_0\rangle=\langle {\overline S} \rangle$ 
and $\langle N_0^c\rangle=\langle {\overline N^c} \rangle$ as 
%%%%%%%%%%%%%%%%%%%%%%%%%%%%%%%%%%%%%%%%%%%%%
\be
   G = SU(6) \times SU(2)_R 
     \buildrel \langle S_0 \rangle \over \longrightarrow 
             SU(4)_{\rm PS} \times SU(2)_L \times SU(2)_R  
     \buildrel \langle N^c_0 \rangle \over \longrightarrow 
     G_{SM}, 
\ee
%%%%%%%%%%%%%%%%%%%%%%%%%%%%%%%%%%%%%%%%%%%%%
where $SU(4)_{\rm PS}$ represents the Pati-Salam 
$SU(4)$\cite{Pati}. 
Hereafter it is supposed that the symmetry breaking scales 
are roughly $\langle S_0 \rangle = 10^{17 \sim 18}$GeV and 
$\langle N^c_0 \rangle = 10^{15 \sim 17}$GeV. 
Gauge invariant trilinear couplings in the superpotential 
$W$ are of the forms 
%%%%%%%%%%%%%%%%%%%%%%%%%%%%%%%%%%%%%%%%%%%%%%
\bea
    (\phi ({\bf 15},{\bf 1}))^3 & = & QQg + Qg^cL + g^cgS, 
\label{W1}
                                                            \\
    \phi ({\bf 15},{\bf 1})(\psi (\overline{\bf 6},{\bf 2}))^2 & 
            = & QH_dD^c + QH_uU^c + LH_dE^c  + LH_uN^c 
                                            \nonumber \\ 
             {}& & \qquad   + SH_uH_d + 
                     gN^cD^c + gE^cU^c + g^cU^cD^c.
\label{W2}
\eea 
%%%%%%%%%%%%%%%%%%%%%%%%%%%%%%%%%%%%%%%%%%%%%%

{}From the viewpoint of the string unification theory, 
it is reasonable that the hierarchical structure 
of Yukawa couplings is attributable to some kind of 
the flavor symmetry at the string scale $M_S$. 
If there exists a flavor symmetry such as $U(1)_F$ 
in the theory, 
it is natural that the Froggatt-Nielsen mechanism is 
at work for the interactions\cite{F-N}. 
For instance, the effective Yukawa interactions for 
up-type quarks are of the form\cite{Matsu1} 
%%%%%%%%%%%%%%%%%%%%%%%%%%%%%%
\be
  M_{ij} \, Q_i U^c_j H_{u0} 
\label{QUH}
\ee
%%%%%%%%%%%%%%%%%%%%%%%%%%%%%%
with 
%%%%%%%%%%%%%%%%%%%%%%%%%%%%%%%%%%%%%%%%%%%%%%%%%%%
\be
    M_{ij} = m_{ij} \left( \frac{\langle X \rangle}
                              {M_S}\right)^{b_{ij}} 
           = m_{ij} \, x^{b_{ij}}, 
\label{Mij}
\ee
%%%%%%%%%%%%%%%%%%%%%%%%%%%%%%%%%%%%%%%%%%%%%%%%%%%
where the subscripts $i$ and $j$ stand for the generation 
indices and the coupling constants $m_{ij}$'s are assumed 
to be ${\cal O}(1)$ with rank\,$m_{ij}=3$. 
$X \equiv (S_0{\bar S})/M_S$ is singlet with a nonzero flavor 
$U(1)_F$ charge and $x \equiv \langle X \rangle /M_S < 1$. 
The exponents $b_{ij}$ are some non-negative integers which 
are settled by the flavor symmetry\footnote
{See Ref.\cite{Matsu1} for the detail realization of the model. 
Here we only give the essence of the model.}. 
The hierarchical mass matrix is derived by assigning appropriate 
flavor charges to the matter fields.

Generally speaking, the mixing occurs not only among 
three generations of low energy matter fields(quarks and letons) 
but also beyond generations. 
Below the scale $\langle N_0^c \rangle$ there appear both 
$D^c$-$g^c$ mixings and $L$-$H_d$ mixings. 
On the other hand, $U^c$ has no state-mixings 
beyond the generation mixing. 
This situation is of great importance to understand the 
characteristic features of $V_{CKM}$ in the present model. 
An early attempt of explaining the CKM matrix via $D^c$-$g^c$ mixings 
has been made in Ref.\cite{Hisano}, in which a SUSY $SO(10)$ model 
was taken.

{}From Eq.(\ref{Mij}) we have the up-quark mass matrix 
%%%%%%%%%%%%%%%%%%%%%%%%%%%%%%%%%%%%%%%%%%%%%%%%%%%%%%%
\be
M = \left(
  \begin{array}{ccc}
 m_{11}x^{\a_1+\b_1} & m_{12}x^{\a_1+\b_2} & m_{13}x^{\a_1} \\
 m_{21}x^{\a_2+\b_1} & m_{22}x^{\a_2+\b_2} & m_{23}x^{\a_2} \\
 m_{31}x^{\b_1}      & m_{32}x^{\b_2}      & m_{33} 
  \end{array}
  \right).
\label{Mu}
\ee
%%%%%%%%%%%%%%%%%%%%%%%%%%%%%%%%%%%%%%%%%%%%%%%%%%%%%%%
The exponents $\a_i$ and $\b_i$ are determined according 
as the flavor $U(1)_F$ charges of matter fields 
and are  assumed to satisfy the relations 
$\a_1>\a_2>\a_3 = 0$ and $\b_1>\b_2>\b_3 = 0$. 
This matrix is diagonalized by the bi-unitary transformation 
as 
%%%%%%%%%%%%%%%%%%%%%%%%%%%%%%%%%%%%%%%%%%%
\be
  M^{diag}={\cal V}_u^{-1}M{\cal U}_u .
\ee
%%%%%%%%%%%%%%%%%%%%%%%%%%%%%%%%%%%%%%%%%%%
Using the perturbative expansion we can obtain the eigenvalues of 
the matrix $M$, which are written in light order as 
%%%%%%%%%%%%%%%%%%%%%%%%%%%%%%%%%%%%%%%%%%%%%%%%%%%%%%%%%%
\be
   m_u \simeq x^{\a_1+\b_1} \left|
        \frac{\det M_0}{\Delta(M_0)_{11}}\right|,\quad
   m_c \simeq x^{\a_2+\b_2}\left| 
        \frac{\Delta(M_0)_{11}}{m_{33}}\right|,\quad
   m_t \simeq \left| m_{33} \right| .
\label{Umass}
\ee
%%%%%%%%%%%%%%%%%%%%%%%%%%%%%%%%%%%%%%%%%%%%%%%%%%%%%%%%%%
Here the matrix $(M_0)_{ij}$ means $m_{ij}$ in Eq.(\ref{Mu}) and 
$\Delta(M_0)_{ij}$ is the cofactor of $(ij)$ element of 
the matrix $M_0$. 
The unitary matrix ${\cal V}_u$ becomes 
%%%%%%%%%%%%%%%%%%%%%%%%%%%%%%%%%%%%%%%%%%%%%%%%%%%%%%%%%%%%%%%%
\be
   {\cal V}_u \simeq 
      \left(
      \begin{array}{ccc}
        1 - {\cal O}(x^{2(\a_1-\a_2)})  &  
         -x^{\a_1-\a_2}\left(\frac{\overline{m}_{21}}
                          {\overline{m}_{11}}\right)^*  &  
            x^{\a_1}\frac{m_{13}}{m_{33}} \\
        x^{\a_1-\a_2}\frac{\overline{m}_{21}}{\overline{m}_{11}}  & 
          1 - {\cal O}(x^{2(\a_1-\a_2)})  & 
            x^{\a_2}\frac{m_{23}}{m_{33}} \\
        x^{\a_1}\frac{\overline{m}_{31}}{\overline{m}_{11}} & 
          -x^{\a_2}\left(\frac{m_{23}}{m_{33}}\right)^*  & 
            1 - {\cal O}(x^{2\a_2})
      \end{array}
      \right) 
\label{Vu}
\ee
%%%%%%%%%%%%%%%%%%%%%%%%%%%%%%%%%%%%%%%%%%%%%%%%%%%%%%%%%%%%%%%%
with
%%%%%%%%%%%%%%%%%%%%%%%%%%%%%%%%%%%
\be
   {\overline m}_{ij} \equiv 
       ({M_0^\dagger}^{-1})_{ij} 
         = \left(\frac{\Delta(M_0)_{ij}}{\det M_0}\right)^*.
\ee
%%%%%%%%%%%%%%%%%%%%%%%%%%%%%%%%%%%
Note that the 3rd column of ${\cal V}_u$ is 
proportional to the 3rd column vector ${\vec M}_3$ of $M$ 
and that the 1st  column  of ${\cal V}_u$ is proportional to 
the 1st column vector ${\vec {\bar M}}_1$ of $(M^\dagger)^{-1}$,
which is proportional to the outer product 
$({\vec M}_2 \times {\vec M}_3)^*$
\footnote{In the previous studies(\cite{Matsu1,Matsu2}) 
the same calculation for ${\cal V}_u$ has been carried out 
up to the ${\cal O}(1)$ factors $m_{ij}$ and ${\bar m}_{ij}$.}.
Another unitary matrix ${\cal U}_u$ is obtained by the 
replacement $m_{ij}\ra m_{ji}^*$ and $\a_i\ra\b_i$ 
in Eq.(\ref{Vu}) for ${\cal V}_u$.

\vskip 1cm
We now proceed to study the down-type quark mass matrix. 
Due to $D^c$-$g^c$ mixings the down-type quark mass matrix 
is expressed in terms of the $6 \times 6$ matrix 
%%%%%%%%%%%%%%%%%%%%%%%%%%%%%%%%%%%%%%%%%%%%%%%%%%%%
\be 
\begin{array}{r@{}l} 
   \vphantom{\bigg(}   &  \begin{array}{ccc} 
          \quad \,  g^c   &  \quad  D^c  &  
        \end{array}  \\ 
\widehat{M}_d = 
   \begin{array}{l} 
        g   \\  D  \\ 
   \end{array} 
     & 
\left( 
  \begin{array}{cc} 
    y_S Z   &     y_N  M  \\
      0     &  \rho_d  M 
  \end{array} 
\right). 
\end{array} 
\label{Md} 
\ee 
%%%%%%%%%%%%%%%%%%%%%%%%%%%%%%%%%%%%%%%%%%%%%%%%%%%%
Three nonzero $3 \times 3$ matrices arise from 
the mass terms $Z_{ij}g_ig_j^c\langle S_0 \rangle$, 
$M_{ij}g_iD_j^c\langle N_0^c \rangle$ and 
$M_{ij}Q_iD_j^c\langle H_{d0} \rangle$, 
where 
%%%%%%%%%%%%%%%%%%%%%%%%%%%%%%%%%
\be
   Z_{ij}=(Z_0)_{ij} x^{\a_i + \a_j+\zeta}
                          = z_{ij}x^{\a_i + \a_j+\zeta}
\ee
%%%%%%%%%%%%%%%%%%%%%%%%%%%%%%%%%
with $\zeta \geq 0$ and $ z_{ij}={\cal O}(1)$.
The exponent $\zeta$ comes from the difference in the flavor
$U(1)_F$ charges between the trilinear products $g_3 g_3^c S_0$
and $g_3 D_3^c N_0^c$.
Here we use the notations $y_S$, $y_N$ and $\rho _d$ for 
the VEV's $\langle S_0 \rangle$, $\langle N^c_0 \rangle$ and 
$\langle H_{d0} \rangle$ in units of the string scale $M_S$, 
respectively. 
{}From Eqs.(\ref{W1}) and (\ref{W2}) it is found that 
the matrix $Z$ is symmetric and that the $M$ is 
the same as the up-type quark mass matrix at the unification scale. 
Since each Yukawa coupling undergoes the radiative corrections distinctively, 
in the way of the renormalization group(RG) evolution to low energy region 
the matrices $M$ in $\widehat{M}_d$ deviate gradually from the matrix $M$ 
for up-type quark mass matrix.
Since $\rho _d$ is very small compared to $y_S$ and $y_N$, 
the mass eigenstates for the left-handed light quarks consist 
almost of $D$-components of the quark doublet $Q$. 
On the other hand, the large mixings between $D^c$ and $g^c$ can 
occur depending on the relative magnitude of $y_S Z$ and $y_N M$. 
The matrix $\widehat{M}_d$ is diagonalized by the bi-unitary 
transformation as 
%%%%%%%%%%%%%%%%%%%%%%%%%%%%%%%%%%%%%%%%%%%%%
\be
   \widehat{M}_d^{diag} = \widehat{\cal V}_d^{-1} 
                       \widehat{M}_d \, \widehat{\cal U}_d
\ee
%%%%%%%%%%%%%%%%%%%%%%%%%%%%%%%%%%%%%%%%%%%%%
The unitary matrices are 
%%%%%%%%%%%%%%%%%%%%%%%%%%%%%%%%%%%%%%%
\bea
  \widehat{\cal V}_d  &  \simeq  &  \left(
  \begin{array}{cc}
       {\cal W}_d  &  -\e (A+B)^{-1} B {\cal V}_d \\
       \e B (A+B)^{-1}{\cal W}_d   &  {\cal V}_d 
  \end{array}
  \right), 
\nonumber\\
  \widehat{\cal U}_d  &  \simeq  &  \left(
  \begin{array}{cc}
      y_S Z^\dagger {\cal W}_d {\Lambda}_d^{(0)-1} 
           &  -y_S^{-1} Z^{-1}{\cal V}_d {\Lambda}_d^{(2)} \\ 
      y_N M^\dagger {\cal W}_d {\Lambda}_d^{(0)-1}
           &   y_N^{-1} M^{-1}{\cal V}_d {\Lambda}_d^{(2)}
  \end{array}
  \right),
\eea
%%%%%%%%%%%%%%%%%%%%%%%%%%%%%%%%%%%%%%%
where 
%%%%%%%%%%%%%%%%%%%%%%%%%%%%%%%%%%%%%%%%%%%%
\be
\e \equiv \frac{\rho_d}{y_N}\ll 1, \qquad 
A \equiv  y_S^2 ZZ^\dagger, \qquad 
B \equiv  y_N^2 MM^\dagger. 
\ee
%%%%%%%%%%%%%%%%%%%%%%%%%%%%%%%%%%%%%%%%%%%%
Since $\e$ is a very small number, the calculation is carried out 
by using the perturbative expansion. 
The unitary matrices ${\cal W}_d$ and ${\cal V}_d$ are determined 
such that Hermite matrices $A+B$ and $(A^{-1}+B^{-1})^{-1}$ is 
diagonalized by the unitary transformations as 
%%%%%%%%%%%%%%%%%%%%%%%%%%%%%%%%%%%%%%%%%%%%%%%%%%%%%%%%%%%
\be
(\Lambda_d^{(0)})^2 = {\cal W}_d^{-1} (A+B) {\cal W}_d, 
\qquad 
(\Lambda_d^{(2)})^2 = {\cal V}_d^{-1} (A^{-1}+B^{-1})^{-1} {\cal V}_d. 
\ee
%%%%%%%%%%%%%%%%%%%%%%%%%%%%%%%%%%%%%%%%%%%%%%%%%
$\Lambda_d^{(0)}$ and $\Lambda_d^{(2)}$ represent three 
eigenvalues for heavy modes with GUT scale masses 
and those for light modes corresponding 
to $d$-, $s$- and $b$-quarks, respectively. 
Let us assume that the (1,1) elements of $A^{-1}$ and $B^{-1}$ 
are of the same order.
Since we obtain $(A^{-1})_{11}={\cal O}((y_S\, x^{2\a_1+\zeta})^{-2})$ and 
$(B^{-1})_{11}={\cal O}((y_N \, x^{\a_1+\b_1})^{-2})$, this assumption 
implies $y_S\, x^{\a_1+\zeta} \simeq y_N \, x^{\b_1}$. 
Hereafter we refer this assumption to as large $D^c$-$g^c$ mixings.

Under the assumption of large $D^c$-$g^c$ mixings 
the mass eigenvalues for light quarks are given by 
%%%%%%%%%%%%%%%%%%%%%%%%%%%%%%%%%%%%%%%%%%%%%
\bea
   m_d  &  \simeq  & \frac{x^{\a_1+\b_1}}
          {\sqrt{|{\bar z}_{11}|^2+|{\bar m}_{11}|^2}}, 
                                                \nonumber\\
   m_s  &  \simeq  & x^{\a_2+\b_1} 
            \frac{\sqrt{|{\bar z}_{11}|^2+|{\bar m}_{11}|^2}}
              {\left| \left|
              \begin{array}{cc}
                  {\bar m}_{11} & {\bar z}_{11} \\
                  {\bar m}_{21} & {\bar z}_{21} 
              \end{array}
              \right|\right|},   \\
   m_b  &  \simeq  & x^{\b_1 - \a_1 + \a_2} 
            \frac
            {\left| \det M_0 \cdot \det Z_0 \right| \cdot 
            \left|\left|
            \begin{array}{cc}
                 {\bar m}_{11} & {\bar z}_{11} \\
                 {\bar m}_{21} & {\bar z}_{21} 
            \end{array}
            \right|\right|}
           {\sqrt{ \vert\vert
            \begin{array}{ccc}
                {\vec z}_3 & {\vec m}_2 & {\vec m}_3
            \end{array}
            \vert\vert^2 +  x^{-2\eta} 
            \vert\vert
            \begin{array}{ccc}
              {\vec m}_3  &  {\vec z}_2  &  {\vec z}_3 
            \end{array}
            \vert\vert^2}}, \nonumber 
\eea
%%%%%%%%%%%%%%%%%%%%%%%%%%%%%%%%%%%%%%%%%%%%%
where ${\bar z}_{ij}\equiv (Z_0^{\dagger -1})_{ij}$,  
$\eta \equiv (\a_1-\a_2)-(\b_1-\b_2)$ and 
${\vec m}_i$(${\vec z}_i$) means the $i$-th column vector 
of $M_0$($Z_0$). 
It is noticeable that this hierarchical mass pattern of 
down-type quarks is rather different from that of 
up-type quarks. 
This result is in line with experimental facts. 
The eigenstates are approximately written as 
%%%%%%%%%%%%%%%%%%%%%%%%%%%%%%%%%%%%%%%%%%%%%%%%%%%%%%%%%%%%%%%%%%
\bea
  |d\rangle & \approx & \frac{1}
                {\sqrt{|{\bar z}_{11}|^2+|{\bar m}_{11}|^2}}
                 (-{\bar z}_{11}|g_1^c\rangle 
                       +{\bar m}_{11}|D_1^c\rangle), 
                                        \nonumber  \\
  |s\rangle & \approx & \frac{1}
                {\sqrt{|{\bar z}_{11}|^2+|{\bar m}_{11}|^2}}
                 (-{\bar m}_{11}^*|g_1^c\rangle 
                       -{\bar z}_{11}^*|D_1^c\rangle), \\
  |b\rangle & \approx & \frac{-\vert
                  \begin{array}{ccc}
                     {\vec z}_3 & {\vec m}_2 & {\vec m}_3
                  \end{array}
                  \vert^*|g_2^c\rangle 
                  - x^{-\eta} \vert 
                  \begin{array}{ccc}
                     {\vec m}_3  &  {\vec z}_2  &  {\vec z}_3 
                  \end{array}       
                  \vert^*|D_2^c\rangle} 
              {\sqrt{ \vert\vert
                  \begin{array}{ccc}
                     {\vec z}_3 & {\vec m}_2 & {\vec m}_3
                  \end{array}
                  \vert\vert^2 +  x^{-2\eta} 
                  \vert\vert
                  \begin{array}{ccc}
                     {\vec m}_3  &  {\vec z}_2  &  {\vec z}_3 
                  \end{array}
                  \vert\vert^2}}. \nonumber 
\eea
%%%%%%%%%%%%%%%%%%%%%%%%%%%%%%%%%%%%%%%%%%%%%%%%%%%%%%%%%%%%%%%%%%%
The diagonalization matrix for light $SU(2)_L$-doublet 
down-quark sector is 
%%%%%%%%%%%%%%%%%%%%%%%%%%%%%%%%%%%%%%%%%%%%%%%%%%%%%
\be
  {\cal V}_d  \simeq  \left(
          \begin{array}{ccc}
               1-\O(x^{2(\a_1-\a_2)}) & 
                   -x^{\a_1-\a_2} a_{21}^* & x^{\a_1} a_{13} \\
               x^{\a_1-\a_2} a_{21} & 
                   1-\O(x^{2(\a_1-\a_2)}) & x^{\a_2} a_{23} \\
               x^{\a_1} a_{31} &  -x^{\a_2} a_{23}^* & 
                                1-\O(x^{2\a_2}) 
          \end{array}
          \right), 
\label{Vd}
\ee
%%%%%%%%%%%%%%%%%%%%%%%%%%%%%%%%%%%%%%%%%%%%%%%%%%%%%
where 
%%%%%%%%%%%%%%%%%%%%%%%%%%%%%%%%%%%%%%%%%%%%%%%%%%%%%%%%%%%%%%
\bea
  a_{21} = \frac { {\bar z}_{11}^*{\bar z}_{21}
                        +{\bar m}_{11}^*{\bar m}_{21} }
                 {|{\bar z}_{11}|^2+|{\bar m}_{11}|^2}\,, \qquad &
  a_{13}^* = \frac{
            \left| 
            \begin{array}{cc}
             {\bar m}_{21} & {\bar z}_{21} \\
             {\bar m}_{31} & {\bar z}_{31} 
            \end{array}
            \right| }
            {\left|
            \begin{array}{cc}
             {\bar m}_{11} & {\bar z}_{11} \\
             {\bar m}_{21} & {\bar z}_{21} 
            \end{array}
            \right| }\,,                 \nonumber \\
  a_{31} = \frac { {\bar z}_{11}^*{\bar z}_{31}
                       +{\bar m}_{11}^*{\bar m}_{31} }
                 {|{\bar z}_{11}|^2+|{\bar m}_{11}|^2}\,, \qquad &
  a_{23}^* = -\frac{
              \left| 
            \begin{array}{cc}
             {\bar m}_{11} & {\bar z}_{11} \\
             {\bar m}_{31} & {\bar z}_{31} 
            \end{array}
            \right| }
            {\left|
            \begin{array}{cc}
             {\bar m}_{11} & {\bar z}_{11} \\
             {\bar m}_{21} & {\bar z}_{21} 
            \end{array}
            \right| }\,. 
\label{aij}
\eea
%%%%%%%%%%%%%%%%%%%%%%%%%%%%%%%%%%%%%%%%%%%%%%%%%%%%%%%%%%%%%%
It is worth noting that the 1st column of ${\cal V}_d$ is 
proportional to a linear combination of ${\vec{\bar Z}_1}$ and 
${\vec{\bar M}_1}$ and that the 3rd column is to 
$({\vec{\bar M}_1}\times{\vec{\bar Z}_1})^*$. 
These significant features are obtained if and only if large 
$D^c$-$g^c$ mixings occur.

\vskip 1cm
%\section{CKM matrix and suppression of $V_{ub}$}
We are now in a position to calculate the mixing matrix $V_{CKM}$. 
The $V_{CKM}$ is given by 
%%%%%%%%%%%%%%%%%%%%%%%%%%%%%%%%%%%%%%%%%%%%%%%%%%%%%
\bea
  V_{CKM} & = & {\cal V}_u^{-1} {\cal V}_d  \nonumber \\
       & \simeq & \left(
       \begin{array}{ccc}
           1-{\cal O}(x^{2(\a_1-\a_2)}) & -x^{\a_1-\a_2} c_{21}^*  
                            &  0  \\
           x^{\a_1-\a_2} c_{21} & 1-{\cal O}(x^{2(\a_1-\a_2)}) 
                                        & x^{\a_2} c_{32}^* \\
           x^{\a_1} c_{31} &  x^{\a_2} c_{32} 
                                  & 1-{\cal O}(x^{2\a_2}) 
       \end{array}
       \right), 
\label{Vckm}
\eea
%%%%%%%%%%%%%%%%%%%%%%%%%%%%%%%%%%%%%%%%%%%%%%%%%%%%%
at the leading order with 
%%%%%%%%%%%%%%%%%%%%%%%%%%%%%%%%%%%%%%%%%%%%%
\bea
  c_{21} & = & \frac{{\bar z}_{11}^* \cdot 
                \left|
                \begin{array}{cc}
                   {\bar m}_{11} & {\bar z}_{11} \\
                   {\bar m}_{21} & {\bar z}_{21} 
                \end{array}
                \right|}
                {{\bar m}_{11}
                   (|{\bar z}_{11}|^2+|{\bar m}_{11}|^2)}\,, 
                        \nonumber   \\
  c_{32} & = & \frac{{\bar m}_{11} \cdot 
                \left|
                \begin{array}{ccc}
                   {\vec m}_3 & {\vec z}_2 & {\vec z}_3 
                \end{array}
                \right|^*}
                {m_{33}^* \cdot (\det Z_0)^* \cdot
                \left|
                \begin{array}{cc}
                    {\bar m}_{11} & {\bar z}_{11} \\
                    {\bar m}_{21} & {\bar z}_{21}
                \end{array}
                \right|}\,,        \\
  c_{31} & = & c_{32} \cdot c_{21} .  \nonumber 
\eea
%%%%%%%%%%%%%%%%%%%%%%%%%%%%%%%%%%%%%%%%%%%%%
If we take the parameterization $x^{\a_1}=\l^3$, 
$x^{\a_1 - \a_2}=\l$ and $c_{21}, c_{32}=\O(1)$, 
the $V_{CKM}$ in Eq.(\ref{Vckm}) is in accord with experimental 
facts. 
It should be emphasizing that the element (1,3) of $V_{CKM}$, 
i.e., $V_{ub}$ vanishes at the leading order. 
This is due to the fact that the 1st column of ${\cal V}_u$ 
is proportional to ${\vec {\bar M}_1}$ and 
that the 3rd column of the matrix ${\cal V}_d$ is proportional 
to $({\vec{\bar M}_1} \times {\vec{\bar Z}_1})^*$. 
Then we must pick up the next-to-leading term to obtain 
a nonzero value for $V_{ub}$. 
Concretely, we obtain 
%%%%%%%%%%%%%%%%%%%%%%%%%%%%%%%%%%%%%%%%
\be
   V_{ub} \simeq  x^{\a_1+2(\b_1-\b_2)} c_{13}
\label{Vub2}
\ee
%%%%%%%%%%%%%%%%%%%%%%%%%%%%%%%%%%%%%%%%
with 
%%%%%%%%%%%%%%%%%%%%%%%%%%%%%%%%%%%%%%%%%%%%%%%%%
\be
  c_{13} = - \frac{{\bar m}_{12} \cdot 
                \left|
                \begin{array}{ccc}
                   {\vec m}_3 & {\vec z}_2 & {\vec z}_3 
                \end{array}
                \right|}  {|{\bar m}_{11}|^2 
                   \cdot (\det M_0)^* \cdot \det Z_0 \cdot 
                \left|
                \begin{array}{cc}
                    {\bar m}_{11} & {\bar z}_{11} \\
                    {\bar m}_{21} & {\bar z}_{21}
                \end{array}
                \right|^*}\,. 
\ee
%%%%%%%%%%%%%%%%%%%%%%%%%%%%%%%%%%%%%%%%%%%%%%%%%
We obtain the order of $V_{ub}$ to be $\sim\lambda^7$ for the parameterization $x^{\alpha_1}\sim\lambda^3,x^{\alpha_2}\sim\lambda^2, x^{\beta_1}\sim\lambda^4$ and $x^{\beta_2}\sim\lambda^2$, which reasonably reproduce the masses of u and c given in Eq.(\ref{Umass}), with $c_{13}\sim\O(1)$. 
The magnitude of $V_{ub}$ is too small compared to the current experimental value, which suggests $V_{ub}\sim \lambda^4$\cite{PDG} as shown later.
In order to obtain reasonable $V_{ub}$ at low energies we have to take the RG effects 
into account. 
Due to the RG effects the mass matrix $M$ for up-type quarks and those in 
${\widehat M_d}$ for down-type quarks which coincide with each other at the 
unification scale, deviate from each other at low energies. 
If the RG corrections for the leading terms of Yukawa couplings are of $O(10\%)$, 
then the RG corrections for ${\cal V}_u$ and  ${\cal V}_d$ dominate over 
the next-to-leading terms of them at low energies. 
Thus, the value of $V_{ub}$ at low energies becomes large compared to Eq.(\ref{Vub2}) 
but remains small compared to $V_{td}$ and $V_{cb} \cdot V_{us}$.
%%%%%%%%%%%%%%%%%%%%%%%%%%%%%%%%%%%%%%%%%%%%%%%%%%%%%%%%%%%%%%%%%

On the other hand, the (3,1) element, i.e., $V_{td}$ 
has a nonzero value at the leading order. 
The 3rd column of ${\cal V}_u$ is proportional to ${\vec M}_3$. 
The 1st column of the matrix ${\cal V}_d$ is proportional 
to a linear combination of ${\vec{\bar Z}_1}$ and 
${\vec{\bar M}_1}$ as pointed out in Eq.(\ref{aij}). 
${\vec M}_3$ is orthogonal to 
$({\vec{\bar M}_1})^* \propto {\vec M}_2 \times {\vec M}_3$ 
but not to $({\vec{\bar Z}_1})^*$ in general. 
In other words, a nonvanishing leading term of $V_{td}$ 
is a consequence of large $D^c$-$g^c$ mixings, 
in which (1,1) elements of $A^{-1}$ 
and $B^{-1}$ are comparable to each other. 
{}From Eq.(\ref{Vckm}) we obtain the relations 
%%%%%%%%%%%%%%%%%%%%%%%%%%%%%%%%%%%%%%%
\bea
     V_{td} & = & V_{cd} \cdot V_{ts},   \label{Vtd}\\
(0.004\sim 0.013)& & (0.0076\sim 0.0094)  \nonumber \\
   |V_{ub}| & < & |V_{cb} \cdot V_{us}|. \label{Vub}\\
(0.0018\sim 0.0045) & & (0.0078\sim 0.0094)  \nonumber 
\eea
%%%%%%%%%%%%%%%%%%%%%%%%%%%%%%%%%%%%%%%
Current experimental values cited in the 
parentheses\cite{PDG}, 
strongly suggest that these relations are viable. 

\vskip 1cm
%\section{summary}
We studied the $V_{CKM}$ in $SU(6) \times SU(2)_R$ model, 
in which $D^c$-$g^c$ mixings occur in down-type quark sector 
but no such mixings in up-type quark sector. 
Nontrivial $V_{CKM}$ is induced by large $D^c$-$g^c$ mixings, 
which are expressed in terms of Eq.(\ref{Md}). 
In the present model $V_{ub}$ is naturally suppressed compared 
to $V_{td}$. 
Introducing the flavor symmetry and Froggatt-Nielsen mechanism, 
we understood the reason why the $V_{CKM}$ is nearly equal to 
unity but not exactly unity. 
Further we obtain phenomenologically viable relations (\ref{Vtd}) 
and (\ref{Vub}). 
The conditions on $\widehat{M}_d$ for yielding the above relations are 
in order as 
%%%%%%%%%%%%%%%%
\begin{enumerate}
\item
The ($D$, $D^c$)-block matrix $M$ in $\widehat{M}_d$ is 
exactly the same as the up-type quark mass matrix 
 at the unification scale.

\item
There appears null or negligibly small ($D$, $g^c$)-block 
matrix in $\widehat{M}_d$ relative to the ($D$, $D^c$)-block matrix. 

\item
There occur large mixings between $D^c$ and $g^c$ with 
$|\langle N^c_0 \rangle| \gg |\langle H_{d0} \rangle|$. 
\end{enumerate}
%%%%%%%%%%%%%%%
The 1st condition is not satisfied in the $SU(5)$ GUT model 
because $U^c$ and $D^c$ belong to the different irreducible 
representations of $SU(5)$. 
Therefore, $V_{ub}$ is not suppressed relative to $V_{td}$ 
in generic $SU(5)$ GUT model. 
The 1st condition means that $SU(2)_R$ is contained in 
the unification gauge group. 
In $SO(10)$ GUT model the circumstances are obscure 
due to variations of Higgs representations.

In the next step we would like to extend this model to 
lepton sector to interpret the large mixing observed in 
neutrino oscillations\cite{Atmos}. 
In lepton sector there exist mixings between $L$ and $H_d$ 
similar to $D^c$-$g^c$ mixings in quark sector. 
In neutral lepton sector we have to take right-handed 
Majorana mass matrices and the seesaw mechanism into account. 
This study is now in progress\cite{Matsu5}.

%%%%%%%%%%%%%%%%%%%%%%%%%%%%%%%%%%%%%%%%%%%%%%%%%%%%%%%%
\section*{Acknowledgements}
Authors would like to thank valuable discussions at 
the research meeting on neutrino physics held at RIFP 
at Kyoto on 28th, February to 1st, March, 2000. 
One of the authors(T. M) is supported in part by a Grant-in-Aid 
for scientific Research from Ministry of Education, Science, 
Sports and Culture, Japan (No. 10640256).

%%%%% Section bibliography %%%%%%%%%%%%%%%%%%%%%%%%%%%

\end{document}